\documentclass{PoS}

\usepackage{epsfig}

\title{The non-perturbative BRST quartets generated by transverse gluons or
 quarks in Landau gauge}

\ShortTitle{The non-perturbative BRST quartets generated by transverse gluons or
 quarks}

\author{\speaker{Natalia Alkofer}
\\
        Institut f\"ur Physik,
Karl-Franzens-Universit\"at,
Universit\"atsplatz 5,
A-8010 Graz, Austria\\
        E-mail: \email{natalia.alkofer@edu.uni-graz.at}}

\author{Reinhard Alkofer\\
Institut f\"ur Physik,
Karl-Franzens-Universit\"at,
Universit\"atsplatz 5,
A-8010 Graz, Austria\\
E-mail: \email{reinhard.alkofer@uni-graz.at}}

\abstract{
 The BRST quartet mechanism is briefly reviewed. A special emphasis is given to
 the distinction of perturbative versus non-perturbative quartets.
 The field contents of the non-perturbative BRST quartets generated by
 transverse gluons or quarks in Landau gauge are presented. 
 Corresponding truncated Bethe-Salpeter equations for the respective
 first daughter and second parent states are
 derived. It is discussed in which sense these equations provide evidence 
 for the existence of bound states as daughter states in non-perturbative 
 BRST quartets. 
 It is noted that within the scaling  solution of functional approaches
 the infrared divergence of the quark-gluon vertex is exactly the right one to 
 make the respective Bethe-Salpeter equation infrared consistent.
}

\FullConference{The many faces of QCD\\
		November 2-5, 2010\\
		Gent Belgium}

\begin{document}

\section{Motivation}

The gluon propagator of Landau gauge QCD has been shown to be positivity
violating, see {\it e.g.\/} Ref.\ \cite{Bowman:2007du}  and references therein.
This especially implies that the one-gluon-state (treated as a physical state
in perturbation theory) belongs to the states of negative norm in the
indefinite-metric state space of Yang-Mills (YM) theory. As such it can be
identified with a parent state in a BRST quartet whose other members, however,
have to be non-perturbative, {\it i.e.\/} bound, states. In the following we
will identify possible members of this quartet and describe a strategy to
provide evidence for their role in the formalism of covariantly gauge-fixed YM
theory. If successful this may provide a detailed picture of the kinematical
aspects of gluon confinement in the Landau gauge. For the quark propagator the
situation is less clear. Nevertheless, by following the same strategy we want
to contribute to a clarification whether quarks are also positivity violating.

\section{The perturbative BRST quartet mechanism}

The perturbative BRST quartet mechanism is the generalization of the
Gupta-Bleuler mechanism \cite{Gupta:1949rh,Bleuler:1950cy} to YM theories, for 
a concise modern treatment see also \cite{Peskin:1995ev,Weinberg:1996kr}.  The
underlying idea is that the gauge condition 
\begin{equation} 
\partial ^\mu A_\mu =0 
\end{equation}
as formulated in classical physics cannot be elevated consistently to an
operator condition in Quantum Field Theory. The correct treatment is instead to
define within the space of all quantum states of QED a physical subspace which
is then given by  the kernel of the operator $\partial^\mu A_\mu^{(+)}$
constructed from $\partial^\mu A_\mu$ by projection on positive energies. To be
concise: The physical state space contains all states $|\Psi\rangle $  which
fulfill
\begin{equation} 
\partial^\mu A_\mu^{(+)} |\Psi\rangle =0 .
\end{equation}
These physical states contain then the longitudinal and the time-like
photons such that their respective contributions precisely cancel. 
Therefore there is no contribution of unphysical states in the $S$-matrix. 
Due to the Minkowski metric it is unavoidable that in covariant gauges
the time-like photon states are negative-metric states, and the total
state space is an indefinite-metric state space. 

\begin{figure}[th]
\centerline{\psfig{file=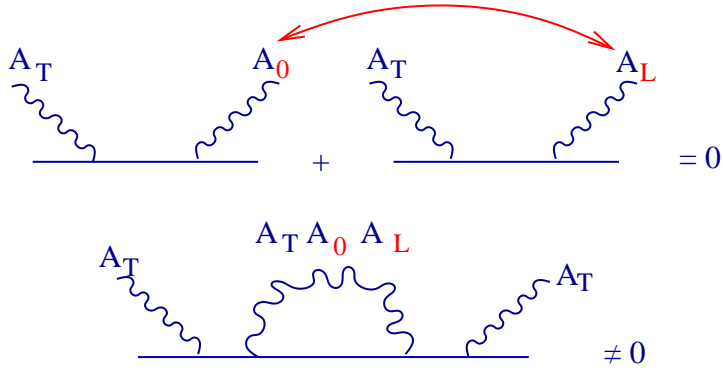,width=100mm}}
\caption{An illustration of the Gupta-Bleuler mechanism in covariantly gauge
fixed QED.
\label{QED}}
\end{figure}

Why then keeping the time-like and the longitudinal photon in the formalism if
they cancel in all physical states? If one includes quantum fluctuations we need
a tool to count them correctly. {\it E.g.\/} in perturbation theory in
non-relativistic quantum mechanics one injects a complete set of states ({\it
i.e.\/} a ``one'') to obtain the correct formulae. The analogue in relativistic
quantum field theory are loops in Feynman diagrams: 
They  describe the quantum fluctuations,
and in order to count correctly one has to inject again a complete set of
states, or phrased otherwise, one has to sum over the propagators of all fields
in the formalism, even the unphysical ones. This way of counting is illustrated
in Fig.~\ref{QED}.

The gauge fields of YM theories (called generically gluons in the following
although the formalism, of course, is valid for all YM theories and not only the
Strong Interactions) are self-interacting. Especially the fact that transverse
gluons may scatter into longitudinal and time-like ones does not allow a
straightforward generalization of the Gupta-Bleuler mechanism. However, on a
purely perturbative level the cancellation mechanism is only slightly more
complicated: Instead of two respective states four do cancel against each other.
In order to describe this so-called quartet cancellation mechanism we discuss
first the elementary BRST quartet \cite{Nakanishi:1990qm}. Within the
Faddeev-Popov quantization of QED one can (by simply ignoring the fact that the
ghosts decouple from the gauge bosons) also formulate the elementary BRST
quartet \cite{Weinberg:1996kr}. This results in, of course, the cancellation
of time-like and longitudinal photons as in the Gupta-Bleuler mechanism. In
non-Abelian gauge theories the elementary BRST quartet takes care of the
cancellation of longitudinal and time-like gluons as well as ghosts and
antighosts in all physical states. Here two remarks are in order: First, the
elementary BRST quartet is also  valid in the limit of gauge coupling $g\to 0$.
Therefore the perturbative BRST quartet mechanism in  YM is only an $m$-fold
duplication of the single cancellation mechanism in QED \cite{Weinberg:1996kr}
with $m$ being the dimension of the adjoint representation of the gauge group. 
Second, due to the nature of the BRST transform one does not directly consider
the longitudinal and time-like gluons but linear superpositions of them, 
the forward, 
resp., backward polarized gluons,  see {\it e.g.} Chapter 16 of
Ref.~\cite{Peskin:1995ev} for a definition of these states.

\begin{figure}[th]
\centerline{\psfig{file=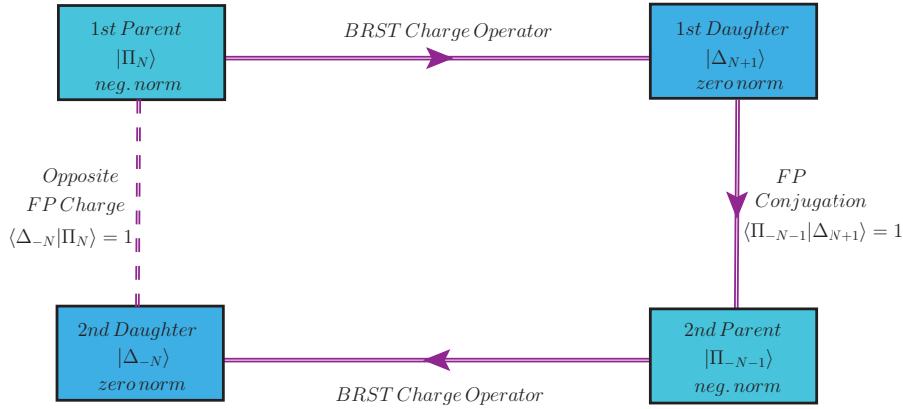,width=120mm}}
\caption{An illustration of the construction of a BRST quartet.
\label{BRSTquartet}}
\end{figure}

The reason for building quartets is related to the nilpotency of the BRST
transformation: Every non-singlet state can then produce only one further state
when the BRST charge operator is applied,
making thus a doublet. It
proves useful to form quartets. This is done such that the Faddeev-Popov charge
conjugated state of the daughter state in this doublet is used as a 2nd
parent state
which under BRST generates the 2nd daughter and thus completes the quartet. The
construction mechanism is illustrated in Fig.\ \ref{BRSTquartet}, and
we will return to it several times in the following. 

To highlight  the nilpotency of the BRST transformation we will work in a
representation with Nakanishi-Lautrup field $B^a$ which becomes on-shell
identical to the  gauge fixing condition, $B^a=({1}/{\xi}) \, \,\partial_\mu
A_\mu^a$ where $\xi$  is the gauge parameter of linear covariant gauges.
To memorize the BRST transformation $\delta_B$ one may picture it  as 
a kind of gauge
transformation with a constant ghost field as parameter:
\begin{equation}
\begin{array}{ll}
\delta_B A^a_\mu \, =\,  \widetilde Z_3 D^{ab}_\mu c^b \, \lambda  \; , 
\quad &  
\delta_B q
\, = \,-  i g t^a \widetilde Z_1 \, c^a \, q \, \lambda \; , \\
\delta_B c^a \, = \, - \, \frac{g}{2} f^{abc} \widetilde Z_1 
 \, c^b c^c \, \, \lambda \; ,
\quad  & \delta_B\bar c^a \, = \, B^a
\, \lambda \; , 
\qquad \qquad 
\delta_B B^a \, = \, 0,\end{array}  
\label{BRST}
\end{equation}
where $D^{ab}_\mu$ is the covariant derivative. The parameter $\lambda$ lives
in the Grassmann algebra of the ghost fields $c^a$ and carries ghost number
$N_{\mbox{\tiny FP}} = -1$. $\widetilde Z_1$ and $\widetilde Z_3$ are the
ghost-gluon-vertex and the ghost wave function renormalization constants.
It has been shown that in Landau gauge $\widetilde Z_1 =1$ \cite{Taylor:1971ff}.

In a next step one follows the construction of Noether's theorem to derive a
 BRST charge operator $Q_B$. 
It generates a ghost number graded algebra on the fields, $\delta_B\Phi = \{
i Q_B, \Phi \}$. Defining the ghost number operator $Q_c$ one obtains the
algebra
\begin{equation} 
Q_B^2 = 0 \; , \quad \left[ iQ_c , Q_B \right] = Q_B \; , 
\quad \left[ Q_c , Q_c \right] = 0 \; .
\label{algebra}
\end{equation}
It is complete in the full and therefore indefinite metric state space of a YM
theory. The BRST cohomology is then constructed as follows: The
semi-definite  physical subspace  $\mbox{Ker}\, Q_B  $ is defined on the basis
of this algebra by those states which are annihilated by the BRST charge $Q_B$,
~$Q_B |\psi \rangle =0$. Since $Q_B^2 =0 $, this subspace contains the space $
\mbox{Im}\, Q_B $ of the so-called daughter states $Q_B |\phi \rangle$
({\it cf.} Fig.\ \ref{BRSTquartet}),
which are
images of their parent states in the indefinite metric  state space.
A physical ({\it i.e.\/} positive-metric) 
Hilbert space is then obtained as the quotient space of equivalence classes:
\begin{equation}
     {\mathcal{H}}(Q_B) = {\mbox{Ker}\, Q_B}/{\mbox{Im}\, Q_B}
        \; .
\end{equation}
This Hilbert space is nothing else than the space of BRST singlets. 
All states are either BRST singlets or belong to quartets,
this exhausts all possibilities \cite{Nakanishi:1990qm}.
Here a remark is in order: Had we required only 
$Q_B |\psi\rangle = 0$
half of these metric partners had been eliminated from  all $S$-matrix elements,
leaving the unpaired daughter states of zero norm 
which do not contribute to any observable, {\it cf.}~Fig.\ \ref{BRSTquartet}.
However, from a mathematical point of view it is more satisfactory to retain
only positive-norm states in the physical state space. Note furthermore that the
parent-daughter states of opposite Faddeev-Popov charge possess non-vanishing
matrix elements (which are usually normalized to one) \cite{Nakanishi:1990qm}.
This elucidates why quartets and not doublets are considered: These
non-vanishing matrix elements are essential in the cancellation mechanism.

As BRST is a symmetry and the BRST charge operator $Q_B$ commutes with the
Hamiltonian the daughter state is degenerate with the parent 
state:\footnote{We thank Dan
Zwanziger for pointing this out to us.}
\begin{equation}
H  |\psi \rangle = E |\psi \rangle  \Rightarrow
 H  Q_B |\psi \rangle =  Q_B H  |\psi \rangle = E Q_B |\psi \rangle .
\end{equation}
And as the Landau gauge  Hamiltonian $H$ is ghost-antighost symmetric 
all members of a BRST quartet are degenerate.

The  elementary quartet consists of the asymptotic states related to the
backward and forward polarized gluons as well as the ghost and the antighost
\cite{Kugo:1979gm,Nakanishi:1990qm}. Hereby one gluon polarization and the
antighost provide the parent states, the orthogonal gluon polarization and the
ghost the daughter states. In all physical states the contribution of this
quartet cancels strictly due to the algebra (\ref{algebra}), a detailed
description is given in Sect.~4.1 of \cite{Nakanishi:1990qm}.
The corresponding construction of perturbative  ``multi-particle'' BRST quartets
follows straightforwardly and is illustrated in Fig.\ \ref{BRSTquartet}.
To fix the notation: We will call the negative norm state of Faddeev-Popov
charge $N$ we start with the 
1st parent $\Pi_N$. Acting with the BRST charge operator $Q_B$ one obtains the 
1st daughter. The Faddeev-Popov charge
reflected state of the 1st daughter provides the 2nd parent. Acting on it with
$Q_B$ provides the 2nd daughter with again has then Faddeev-Popov
charge $N$.

\section{The non-perturbative BRST quartet mechanism}

Within perturbation theory the transverse gluons belong to the Hilbert space
defined by the BRST cohomology. However, this is in open conflict with the
observed confinement of gluons. Therefore it has been conjectured already in the
seventies that the transverse gluons are also part of a BRST quartet
\cite{Kugo:1979gm}. This property is then in turn believed to be an important 
aspect of
gluon confinement \cite{vonSmekal:2000pz}. 
Somewhat later it has been observed
\cite{Oehme:1980ai} that the antiscreening of gluons (which is a very welcome
property as it explains asymptotic freedom) is already at the perturbative level
in conflict with the positivity of the gluon spectral function. As stated above
there is
no doubt any more that the transverse gluons of Landau gauge QCD are positivity
violating, see {\it e.g.} Ref.~\cite{Bowman:2007du}. 

An inspection of  Fig.\ \ref{BRSTquartet} implies 
that ``one-transverse-gluon'' states
are BRST parent states. Their respective daughters, however, cannot be the
elementary ``one-ghost'' states because these are members of the elementary
quartet. From eq.\ (\ref{BRST}) it immediately follows that the
1st daughter state of an ``one-transverse-gluon'' state needs to have 
the field content
 $  \widetilde Z_3 f^{abc} A^c_\mu c^b \, .$
For every ``one-transverse-gluon'' state there should occur exactly one
degenerate 
daughter state. This implies the existence of a ghost-gluon bound state in the
adjoint representation \cite{Alkofer:2011pe}.
 In this sense the resulting BRST quartet is strictly
non-perturbative because the formation of bound states cannot be 
described  with perturbation theory. The Faddev-Popov charge
reflected 2nd parent state is then an antighost-gluon bound state. 
In this context Landau gauge
provides an advantage as compared to general linear covariant gauges: In the
limit $\xi\to0$ the formalism becomes ghost-antighost-symmetric, and thus the
existence of a ghost-gluon bound state implies the occurrence of a degenerate
antighost-gluon bound state with same quantum numbers. Even having then the 2nd
parent, the BRST transformation (\ref{BRST}) leaves then three possibilities
for the 2nd daughter: a ghost-antighost bound state, a ghost-antighost-gluon
bound state, or a bound state of two differently polarized gluons.

Besides the almost trivial observation that, if a BRST quartet is generated by
quarks it can only be a non-perturbative one, containing a ghost-quark bound
state as 1st daughter not much is known about BRST
quartets generated by quarks. It is also unknown whether quarks violate positivity.
Although for light quarks dynamical chiral symmetry breaking (and for heavy
quarks explicit chiral symmetry breaking) determines the infrared behaviour of
the quark propagator the analytic structure of the quark propagator is highly
sensitive to details in the quark-gluon vertex,  see, {\it e.g.\/},
Ref.~\cite{Alkofer:2003jj}. The quark-gluon vertex for light quarks is, on the
other hand, also very strongly influenced by dynamical chiral symmetry breaking
\cite{Skullerud:2003qu,Alkofer:2006gz,Alkofer:2008tt}. The mass generation for
quarks related to chiral symmetry breaking depends strongly on details of
the dynamics. Which mechanism then guarantees that the corresponding bound
states are degenerate with the quark states is completely unknown. We therefore
hope that an investigation of non-perturbative BRST quartets at least partially
will help to resolve these questions.

\section{Properties of ghost-gluon bound states from infrared Landau gauge
YM theory}

By now quite some information on the infrared behaviour of  Landau gauge YM
theory is available. Especially, in the deep infrared
general properties have been deduced by employing functional equations. 
Dyson-Schwinger equation studies have been extended from a previous
analysis of gluon and ghost propagators
\cite{vonSmekal:1997is,Watson:2001yv,Zwanziger:2001kw,Lerche:2002ep,Fischer:2002hna}
to all Yang-Mills vertex functions 
\cite{Alkofer:2004it,Huber:2007kc,Alkofer:2010tq}. 
Functional Renormalization Group Equations allow a further restriction on 
the solution for the Green's functions:
There is one
unique scaling solution with power laws for the Green's functions
\cite{Fischer:2006vf,Fischer:2009tn} and a one-parameter family of solutions,
the so-called decoupling solutions. The latter are infrared trivial
solutions which possess as an endpoint exactly the scaling solution 
characterized by infrared power laws.  Numerical
solutions of the decoupling type (there called ``massive solution'')
have been published in \cite{Aguilar:2008xm,Boucaud:2008ky} and
references therein. A recent detailed description and comparison of
these two types of solutions has been given in Ref.\
\cite{Fischer:2008uz}, see also Refs.
~\cite{Alkofer:2008jy,Huber:2009wh,Szczepaniak:2001rg,Epple:2007ut}.
Most lattice calculations of the gluon propagator favor  a
decoupling solution. However, in Ref.~\cite{Maas:2009se} it has been suggested
that the infrared behaviour of the Green's function may depend on the
non-perturbative completion of the gauge.

The scaling solution
respects BRST symmetry whereas every decoupling solution breaks it
\cite{Fischer:2008uz}, although very likely only softly. 
Being very strictly,  the
analysis as presented below will be only valid if the scaling
solution is a correct one. The situation is, however, not as severe as it 
seems. First, if the conjecture of Ref.~\cite{Maas:2009se} is correct it is
sufficient that only one non-perturbative completion of Landau gauge with
scaling solution exists to make the analysis of Ref.~\cite{Alkofer:2011pe}
well-founded. Second, even if only
decoupling type of solutions were correct an extended BRST-like
nilpotent symmetry is likely to take the role of the BRST symmetry
\cite{Sorella:2009vt}, or the soft  BRST symmetry breaking can be treated as
spontaneous symmetry breaking \cite{Zwanziger:2010iz}, see also
Ref.~\cite{Sorella:2011tu} and references therein, as well as the discussion
below. It is important to realize
that all arguments about infrared dominance of diagrams stay  correct:
The  numerical value of a diagram which is infrared leading in the scaling
solution will be large in a physically acceptable decoupling solution.

All one-particle
irreducible Green's functions in the scaling solution in the simplified case
with only one external spacelike scale $p^2\to 0$ obey a simple power
law. For a function with
$n$ external ghost and antighost as well as $m$ gluon legs one obtains:
\begin{equation}
\Gamma^{n,m}(p^2) \sim (p^2)^{({n-m})\kappa} .
\end{equation}
Hereby the best known value of $\kappa$ is calculated from truncated equations
and is given by $\kappa =0.595$ \cite{Zwanziger:2001kw,Lerche:2002ep}.
The above solution fulfills all functional equations and all
Slavnov-Taylor identities. It verifies the hypothesis of infrared
ghost dominance \cite{Zwanziger:2003cf}.

As  already emphasized gluons violate positivity
\cite{Bowman:2007du,Alkofer:2003jj}. For the scaling solution this can be
immediately deduced from the fact that for this solution the gluon propagator
vanishes at zero virtuality, $p^2=0$, with an exponent $2\kappa -1$. 
It leads to an infrared diverging
ghost propagator with exponent $-\kappa-1$ as well as infrared diverging 
three- and four-gluon vertex functions ($-3\kappa$ and $-4\kappa$,
respectively).
A further important property of the
scaling solution is the infrared trivial behaviour of the ghost-gluon vertex
which is in agreement with general arguments \cite{Taylor:1971ff,Lerche:2002ep}.

The  ghost-gluon bound state is looked for in the ghost-gluon
scattering kernel. To this end we want to truncate this quantity to the
infrared leading term.  We use the MATHEMATICA package DoDSE
\cite{Alkofer:2008nt,Huber:2010ne} to derive the diagrammatic expressions for
the Dyson-Schwinger equation of this four-point function. A diagram-by-diagram
infrared power counting is performed by attributing anomalous infrared
exponents to the  internal legs and vertex functions. A ghost propagator
provides a $-\kappa$, a gluon propagator a $2\kappa$, the vertex functions the
powers cited above. It is somewhat lengthy but straightforward to verify that
in the scaling  solution the infrared exponent of the
ghost-gluon scattering kernel is $-\kappa$.  More important,
the infrared power counting also provides  the infrared leading terms.

With two different fields 
involved there are two distinct possibilities for the Dyson-Schwinger
equation according to which leg one puts the bare vertex. 
Placing the bare vertex to a ghost leg  provides a consistent infrared counting
\cite{Alkofer:2011pe}.

\begin{figure}[th]
\centerline{\psfig{file=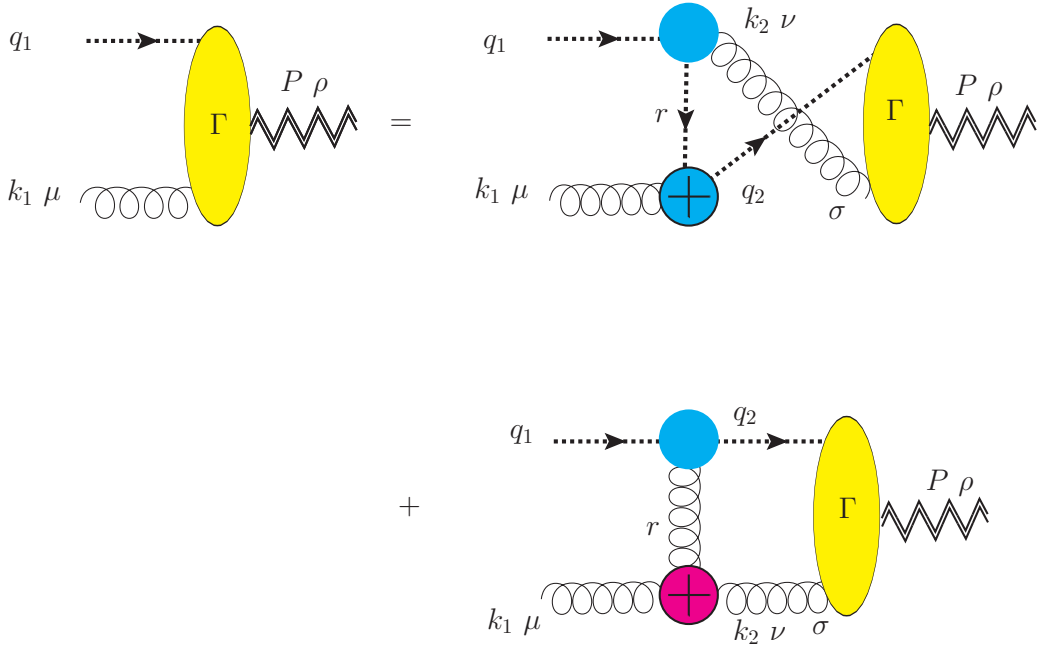,width=140mm}}
\caption{Graphical representation of the gluon-ghost Bethe-Salpeter equation.
Crosses denote dressed vertices.
\label{GhGl}}
\end{figure}

The truncation process 
for the diagrams on the r.h.s to be kept is: It should contain
the one-particle irreducible ghost-ghost-gluon-gluon four-point function and no
$n\ge 5$-point function, it should be infrared leading, and 
the interaction shall take place in the ghost-gluon channel. 
This leaves two diagrams:
One with two ghost and one gluon propagator on
internal lines. This is effectively a ghost exchange. 
And another one with two gluon and one ghost propagator on
internal lines. This is a gluon exchange. Note that this diagram is infrared
leading because in the scaling solution the fully dressed
three-gluon vertex is infrared divergent. 

Assuming the existence of a bound state as well as employing the
usual decomposition of the (ghost-ghost-gluon-gluon)
four-point function into Bethe-Salpeter amplitudes and performing the expansion
around the pole (see {\it e.g.\/} Sect.~6.1 of Ref.~\cite{Alkofer:2000wg}) one
arrives at the Bethe-Salpeter equation depicted in Fig.~\ref{GhGl}.
Using the propagator
parameterizations of {\it e.g.\/} Ref.\ \cite{Alkofer:2003jj}, the
ghost-gluon vertex of Ref.\ \cite{Schleifenbaum:2004id}, and the three-gluon
vertex of Ref.\ \cite{Alkofer:2008dt} one can derive a self-consistent
equation for the corresponding Bethe-Salpeter amplitude containing otherwise
only known quantities \cite{Alkofer:2011pe}.
The decisive property of the kernels of this Bethe-Salpeter equation are: 
For the upper diagram of the r.h.s of Fig.~\ref{GhGl} 
the kernel is well represented by 
\begin{equation}
\alpha^{\mathrm gh}(r^2)/r^2 \qquad {\rm with }  \qquad
\alpha^{\mathrm gh}(r^2)
=\frac{g^2}{4\pi}G^2(r^2) Z(r^2) .
\end{equation}
($G$ and $Z$ are the ghost and gluon renormalization function, respectively.)
For the lower diagram the corresponding expression is 
$\sqrt{\alpha^{\mathrm gh}(r^2)} \sqrt{\alpha^{\mathrm 3g}(r^2)} /r^2$
with $\alpha^{\mathrm 3g}(r^2)$ being proportional to the square of the
three-gluon vertex and
$Z^3$. As the coupling constant derived from the 3-gluon vertex has a smaller
infrared fixed point \cite{Alkofer:2008dt} than the one derived from the
ghost-gluon vertex the upper diagram will be dominant. With 
$\alpha^{\mathrm gh}(0) = 8.92/N_c$ (see {\it e.g.\/} \cite{Alkofer:2002ne} or
Sect.~2.3 of Ref.~\cite{Fischer:2006ub}) it is evident that the kernel of
the ghost-gluon Bethe-Salpeter equation is very strong. As typical strengths for
critical coupling constants are of the order of one (see {\it e.g.\/} 
Ref.~\cite{Alkofer:2000wg}) one may even speculate whether the kernel of this 
Bethe-Salpeter equation provides evidence for a dynamical breaking of BRST
symmetry. A very welcome side effect would be the related Goldstone nature of
the bound state guaranteeing masslessness.

\section{On the quark-gluon bound state equation}

The scaling solution for the YM Green's functions leads to dynamical chiral
symmetry breaking in the quark sector \cite{Alkofer:2008tt}. The quark
propagator is then infrared finite. The twelve possible Dirac tensor structures
of the quark-gluon vertex are then all infrared divergent with an infrared
exponent $-\kappa -1/2$. The same infrared divergence results for vanishing
gluon momentum, and  this leads to an $1/k^4$ behaviour of the kernel in the
four-quark function, $k$ being the momentum exchange. This is indicative of a
linearly rising  potential between static quarks, and thus quark confinement.
Furthermore, the Slavnov-Taylor identities require that the
ghost-ghost-quark-quark scattering kernel is infrared trivial, see Sect.~3.9 in
Ref.~\cite{Alkofer:2008tt}. 

\begin{figure}[bh]
\centerline{\psfig{file=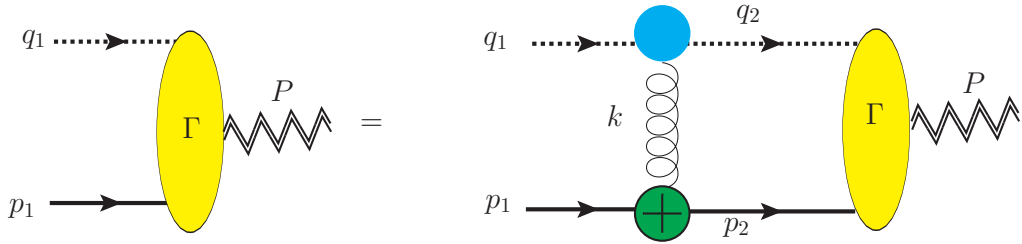,width=140mm}}
\caption{Graphical representation of the quark-ghost Bethe-Salpeter equation.
\label{GhQu}}
\end{figure}

As in the ghost-gluon case one has two choices for the Dyson-Schwinger
equation for the quark-ghost scattering kernel 
according to which leg one puts the bare vertex.
Choosing a ghost leg to place the bare vertex is the infrared consistent 
choice \cite{Alkofer:2011pe}.  Using
the same truncation requirements and the same derivation of the Bethe-Salpeter
equation as in the previous subsection one arrives at the equation depicted in
Fig.~\ref{GhQu}. This equation is in full agreement with the infrared analysis
of the scaling solution, {\it i.e.\/} it is a valid bound state equation, and
in its kernel the infrared exponent $\kappa$ cancels.

Furthermore, this kernel is well approximated  by
$\sqrt{\alpha^{\mathrm gh}(k^2)} \sqrt{\alpha^{\mathrm q-gl}(k^2)} /k^2$
where $\alpha^{\mathrm gh}(k^2)$ is defined above and $\alpha^{\mathrm
q-gl}(k^2)$ is proportional to the square of the quark-gluon vertex and $Z$.
As $\alpha^{\mathrm q-gl}(k^2) \propto 1/k^2$ the above remarks of the
super-criticality of the kernel equally apply. 

\section{Conclusions and outlook}

In these notes we briefly reviewed the concept of BRST quartets, and we emphasized
the different roles of the perturbative and non-perturbative BRST quartets. We
have discussed a possibility how the non-perturbative BRST quartets generated by
transverse gluons and quarks can be studied quantitatively.

To complete this project many open questions still needs to be answered:
What are the bound states representing the respective 2nd daughters? Is BRST
spontaneously broken? Are there associated Goldstone bosons or fermions? Can a
solution of the homogeneous or inhomogeneous Bethe-Salpeter
equation  provide information on the positivity or positivity violation for
quarks? And what is then the relation to quark confinement?

\section*{Acknowledgments}
We thank the organizers of  
this workshop
for all their efforts which made this highly interesting workshop possible. We
are grateful to many of the participants for interesting discussions.

\end{document}